        \documentstyle[11pt]{article}
	\newcommand{\al}{\alpha}
        \newcommand{\del}{\delta}
        \newcommand{\eps}{\epsilon}
        
	\newcommand{\sig}{\sigma}
        \newcommand{\om}{\omega}
	\newcommand{\PHI}{{\it \Phi}}
	\newcommand{\PSI}{{\it \Psi}}
        \newcommand{\Gam}{{\it \Gamma}}
        \newcommand{\Sig}{{\it \Sigma}}

        \newcommand{\EQNO}[1]{\setcounter{equation}{0}
                              \def\theequation{{#1}.\arabic{equation}}}
        \newcommand{\SECT}[2]{\EQNO{{#1}}
                \vspace{4.5ex}
                \begin{center}
                {{#1}. {\bf {#2}}}
                \end{center}}
        \newcommand{\be}{\begin{equation}}
        \newcommand{\ee}{\end{equation}}
        \newcommand{\bea}{\begin{eqnarray}}
        \newcommand{\eea}{\end{eqnarray}}
        \newcommand{\nno}{\nonumber \\}
        \newcommand{\eqlb}[1]{\label{eq: {#1}}}
        \newcommand{\eqrf}[1]{($\!\!$~\ref{eq: {#1}})}
        \newcommand{\sep}[1]{\!\!\!\! &{#1}& \!\!\!\! }
        \newcommand{\eq}{\sep{=}}
        \newcommand{\vc}{\sep{ }}

	\newcommand{\ind}{{\rm ind}\/}
	\newcommand{\grad}{{\rm grad}\;}
        \newcommand{\e}{e}
        \newcommand{\dr}{d}
        \newcommand{\pdr}{{\partial}}
        \newcommand{\inv}[1]{{\textstyle\frac{1}{#1}}}
        \newcommand{\hf}{\inv{2}}
	\newcommand{\D}{{\cal D}}

         \font\BBB=msym10 at 11 pt

        \newcommand{\RE}{{\mbox{\BBB{R}}}}
	\newcommand{\gt}{\tilde{g}}
	\newcommand{\pdt}{\tilde{\pdr}}
	\newcommand{\pdb}{\bar{\pdr}}
	\newcommand{\nabb}{\bar{\nabla}}
	
	\newcommand{\spdo}[2]{{\textstyle\frac{\pdr {#1}}{\pdr {#2}}}}
	\newcommand{\fdo}[2]{\frac{\del {#1}}{\del {#2}}}
	\newcommand{\sfdo}[1]{{\del/\del {#1}}}
	\newcommand{\chr}[2]{\Gam^{\;{#1}}_{{#2}}}
	\newcommand{\chrp}[2]{\Gam'^{\;{#1}}_{{#2}}}
	\newcommand{\chrt}[2]{\tilde{\Gam}^{\;{#1}}_{{#2}}}
	\newcommand{\Chr}{Christoffel }
	\newcommand{\mq}{Mathai-Quillen }
	\newcommand{\diff}{{\mbox{Diff}}}
	\newcommand{\ev}{{\mbox{ev}}}
	\newcommand{\sect}{\Gam(\Sig,X)}
	\newcommand{\map}{{\mbox{Map}}(\Sig,M)}
	\newcommand{\psb}{\bar{\psi}}
	\newcommand{\eab}{\eps^{\;\;\beta}_\al}
	\newcommand{\lag}{{\cal L}}
	\newcommand{\E}{{\cal E}}
	\newcommand{\EA}{{\cal E}^{01}}
	\newcommand{\M}{{\cal M}}
	\newcommand{\R}{{\cal R}}
	\newcommand{\rt}{T(X/\Sig)}

\begin{document}
$\!\,{}$

	\vspace{-5ex}

	\begin{flushright}
{\tt hep-th/9406103}\\
June, 1994
	\end{flushright}

        \begin{center}
        {\LARGE\bf On the Mathai-Quillen Formalism

	\vspace{1ex}
	of Topological Sigma Models}

        \vspace{3ex}

        {Siye Wu\footnote{e-mail address: {\tt sw@shire.math.columbia.edu}}}\\
	\vspace{1ex}
	{\small Department of Mathematics\\
	Columbia University\\
	New York, NY 10027}
        \end{center}

	\vspace{2ex}

	\begin{quote}
{\small $\quad$ We present a \mq interpretation of topological sigma models.
The key to the construction is a natural connection in a suitable
infinite dimensional vector bundle over the space of maps from a
Riemann surface (the world sheet) to an almost complex manifold (the target).
We show that the covariant derivative of the section defined by the
differential operator that appears in the equation for pseudo-holomorphic
curves is precisely the linearization of the operator itself.
We also discuss the \mq formalism of gauged topological sigma models.}
\end{quote}

        \SECT{1}{Introduction}

The theory of pseudo-holomorphic curves has many successful applications to
symplectic geometry since it was introduced by Gromov [\ref{g}].
(For recent surveys, see [\ref{mdal}] and references therein.)
In the topological sigma model [\ref{sig}], one of the several topological
field theories proposed by Witten, functional integrals are localized to the
moduli space of pseudo-holomorphic curves in the (exact) semi-classical limit,
the correlation functions are Donaldson-type invariants in Gromov's theory
and the space of quantum ground states is the Floer group.
These phenomena, in this and other topological field theories, can be
understood most naturally when the field theories are based on infinite
dimensional versions of the \mq construction [\ref{mq}].
(See [\ref{blau}] for a review.)
For example, four dimensional topological Yang-Mills theory [\ref{tym}],
which is related to the works Donaldson and Floer, does have a \mq
interpretation [\ref{aj}].
In this paper, we present the case of topological sigma model and its gauged
versions in the same spirit.

The paper is organized as follows.
In Section 2, we review the \mq construction and its appearance in the
infinite dimensional setting of loop spaces. The relation with the BRST
algebra and Lagrangian in supersymmetric quantum mechanics [\ref{sqm}]
will provide a guidance for the remaining sections.
In Section 3, we consider the space of maps $\map$ from a Riemann surface
$\Sig$ to an almost complex manifold $M$ and an infinite dimensional vector
bundle $\E^{01}$ over $\map$, whose fiber over a map $u$ is the space of
anti-holomorphic sections of the bundle $u^*TM\otimes T^*\Sig$ over $\Sig$.
We show that there is a natural connection $\nabla^{01}$ on $\E^{01}$.
Moreover, the covariant derivative of the section $u\mapsto\pdb_Ju$ along
a tangent vector $\PHI$ of $\map$ is the action of a first order partial
differential operator on $\PHI$ which linearize the section itself.
This provides the \mq interpretation of the topological sigma model.
Section 4 is about its gauged versions.
We replace the maps to the target space in the previous model by sections
of a fibration of Riemannian manifolds.
We find however that the connection of the infinite dimensional
vector bundle in this generalized setting preserves the linear metric
in the fibers only when the parallel transport of the finite dimensional
fibered space generates isometries among the Riemannian fibers.
In this special case, the topological sigma model is coupled to gauge fields
in the usual sense.
Finally, a systematic treatment of the notions of manifold, connection and
curvature in infinite dimensional settings can be found in [\ref{k}].

	\SECT{2}{\mq Formalism and Supersymmetric Quantum Mechanics}

We first recall the basic notion of the \mq construction in
finite dimensional settings.
Consider a vector bundle $E$ of rank $m$ associated to a principal bundle $P$
over a compact base manifold of dimension $n$.
Let $\{x^i\}$ be local coordinates on $M$ and $\{\xi^a\}$, the linear
coordinates on the fiber $F$. Choose a metric $g_{ij}$ on $M$ and a linear
metric $h_{ab}$ in the fiber to raise and lower indices.
Given a connection $\nabla$ on $E$ (compatible with the metric $h_{ab}$),
the Euler class $e(E)$ of the bundle $E$ is the Pfaffian of the curvature
2-form $R^a_{\;\;b}$ and can be written in terms of a fermionic integral.
More generally, let $\rho$ be a fermionic variable in $F$, then
	\be\eqlb{thom}
u_\nabla(E)=\inv{(2\pi)^m}\,\e^{-\frac{|\xi|^2}{2}}
	    \int\dr\rho\,\e^{i\nabla\xi^a\rho_a+\hf\rho_aR^{ab}\rho_b}
	\ee
is a basic form on $P\times F$ and can be regarded as
a representative of the Thom class on $E$ [\ref{mq}].
For any section $s\colon M\to E$, the pull-back
$e_{s,\nabla}=s^*u_\nabla(E)$ is given by the right hand side of \eqrf{thom}
after replacing $\xi^a$ by $s^a$.
The de Rham class of this $m$-form on $M$ does not depend on the choice of
the connection or the section, and is equal to the Euler class $e(E)$
of the bundle $E$.
If $m=n$, we can integrate $e_{s,\nabla}$ over $M$; this gives the Euler
number $\chi(E)$.
Introducing another fermionic variable $\chi$ in $TM$, then
	\be\eqlb{char}
\chi(E)=\inv{(2\pi)^m}\int\dr x\dr\chi\dr\rho\,
        \e^{-\frac{|s(x)|^2}{2}+i\nabla_is^a(x)\chi^i\rho_a
            +\inv{4}R^{ab}_{\;\;\;ij}\chi^i\chi^j\rho_a\rho_b},
	\ee
which resembles the partition function of a supersymmetric system.
If $m<n$, we have to insert differential forms appropriate degree
in the integrand to get non-zero numbers.
Physically, this amounts to the calculation of the expectation value
of an observable $O=O_{i_1\dots i_{m-n}}(x)\chi^{i_1}\dots\chi^{i_{m-n}}$.
In infinite dimensional cases, the Euler class so defined formally may depend
of the choice of the section [\ref{aj}].
Moreover, the insertion of differential forms or observables is possible
only when the difference $n-m$ is finite, or when the zero locus of the
section is finite dimensional.

Now consider a case in which the base manifold is the (infinite dimensional)
loop space $LM=\mbox{Map}(S^1,M)$ of a compact Riemannian manifold $M$
with metric $g_{ij}$ and the vector bundle is its tangent bundle $T(LM)$.
A tangent vector at a loop $u(t)$ is a section of the pull-back bundle
$u^*TM$ over the circle $S^1$.
Choosing the functional-derivative operators $\sfdo{u^i(t)}$ as a
basis of $T(LM)$, a tangent vector field on $LM$ is locally
	\be\eqlb{vect}
\PHI=\int_{S^1}\dr t\,\PHI^i(u,t)\fdo{}{u^i(t)},
	\ee
or is simply denoted by $\PHI^i(u,t)$.
$LM$ is equipped with an induced metric
	\be
g\left(\fdo{}{u^i(s)},\fdo{}{u^j(t)}\right)=g_{ij}(u(t))\,\del(s-t).
	\ee
So the \Chr symbols and the Riemann curvature of $LM$ are equal to
those of $M$ up to factors of delta functions.
For example, the covariant derivative of a vector field $\PHI^i(u,t)$ is
	\be
\nabla_\sfdo{u^i(s)}\PHI^k(u,t)=
\fdo{\PHI^k(u,t)}{u^i(s)}+\chr{k}{ij}(u(t))\,\del(s-t)\PHI^j(u,t).
	\ee
For each Morse function $W$ on $M$, there is a natural tangent vector field
$\dot{u}+\grad W$ on $LM$, whose components are
${\dr u^i(t)}/{\dr t}+W^{,i}(u(t))$.
The covariant derivative of $\dot{u}$ along a base vector is
	\be\eqlb{11}
\nabla_\sfdo{u^i(s)}\frac{\dr u^k(t)}{\dr t}=\left[-\del^k_i\frac{\dr}{\dr s}
+\chr{k}{ij}\frac{\dr u^j(t)}{\dr t}\right]\del(s-t).
	\ee
Hence for any tangent vector $\PHI$ of $LM$,
	\be\eqlb{12}
\nabla_\PHI[\dot{u}^k(t)+W^{,k}(u(t))]=D_t\PHI^k+\PHI^iW^{,k}_{\;\;\;;i},
	\ee
where $D_t\PHI^k=\;\dot{\!\!\PHI}^k+\chr{k}{ij}\dot{u}^i\PHI^j$ is the
covariant derivative given by the pull-back connection in $u^*TM$.

We now compare the \mq formalism of $T(LM)$ with supersymmetric quantum
mechanics, of which the fundamental variables are a bosonic loop $u^i(t)$,
two fermionic fields $\psi^i(t)$ and $\psb^i(t)$ in the tangent space,
and a (bosonic) multiplier field $B_i(t)$, with ghost numbers
$0$, $1$, $-1$, $0$ respectively.
The BRST algebra is (see for example [\ref{bs}, \ref{blau}])
	\bea
\vc\del u^i=i\psi^i, \quad\quad\quad\quad \del\phi^i=0,		\\
\vc\del\psb_i=B_i-i\chr{k}{ij}\psi_k\psi^j,		\eqlb{psb}\\
\vc\del B_i=i\chr{k}{ij}B_k\phi^j-\hf R^j_{\;\;ikl}\psb_i\psi^k\psi^l.\eqlb{B}
	\eea
The Lagrangian is
	\bea
\lag\eq\del(\psb_i(\dot{u}^i+W^{,i})-\hf\psb_ig^{ij}B_j)		\nno
\eq-\hf B_iB^i+B_i(\dot{u}^i+W^{,i})-i\psb_i(D_t\psi^i+\psi^jW^{,i}_{\;\;\;;j})
   +\inv{4}R^{ij}_{\;\;\;kl}\psb_i\psb_j\psi^k\psi^l.
	\eea
After eliminating $B_i$ using the equation of motion $B^i=\dot{u}^i+W^{,i}$,
we have
	\be\eqlb{lag1}
\lag=\hf g_{ij}(\dot{u}^i+W^{,i})(\dot{u}^j+W^{,j})
     -i\psb_i(D_t\psi^i+\psi^jW^{,i}_{\;\;\;;j})
     +\inv{4}R^{ij}_{\;\;\;kl}\psb_i\psb_j\psi^k\psi^l.
	\ee
Both the BRST algebra and the Lagrangian are related to the \mq
formalism of $T(LM)$.
First, the non-covariant looking terms in \eqrf{psb} and \eqrf{B} are
determined by the term in \eqrf{11} proportional to $\dr u^j(t)/\dr t$,
which is replaced by $\psb^j$.
$\del^2\psb=0$ requires that the additional term in \eqrf{B} is proportional
to $\nabla^2\dot{u}$, i.e., the curvature of the infinite dimensional bundle.
That $\del^2B_i=0$ is guaranteed by the (differential) Bianchi identity.
Secondly, in light of \eqrf{12}, the action $\int_{S^1}\dr t\,\lag$
agrees completely with the exponent in \eqrf{char} with the section
$s=\dot{u}+\grad W$ and with $u(t)$, $\psi(t)$ and $\psb(t)$ playing the role
of $x$, $\chi$, $\rho$, respectively.
Hence the partition function
	\be
Z=\int\D u\D\psi\D\psb\;\e^{-\int\dr t\,\lag(u,\psi,\psb)}
	\ee
is formally the Euler characteristic of $LM$ regularized by the
section $\dot{u}+\grad W$.

Let $F_u$ be the differential operator acting on $\PHI$ on the
right hand side of \eqrf{12}, i.e., $F_u\PHI=D_t\PHI+\nabla_\PHI(\grad W)$.
Mathematically, the above \mq interpretation is based on two facts.
First $F_u\PHI=0$ is precisely the linearization of the instanton equation
	\be\eqlb{instanton}
\dot{u}+\grad W(u)=0.
	\ee
Secondly, due to the BRST algebra defined above, the same linearization
also appears as the fermionic kinetic term in the Lagrangian \eqrf{lag1}.
Furthermore, $F_u$ determines the dimension of the space of solutions
of \eqrf{instanton}.
If $u$ is a map from $\RE$ to $M$ satisfying
	\be\eqlb{asymp}
\lim_{t\to-\infty}u(t)=y,\quad\quad\lim_{t\to+\infty}u(t)=x,
	\ee
where $x$ and $y$ are two (isolated) critical points of $W$ with Morse
indices $\ind(x)$ and $\ind(y)$ respectively, then $F_u$ is a Fredholm
operator whose index is (see for example [\ref{s}])
	\be
\ind(F_u)=\ind(y)-\ind(x).
	\ee
Let $\M(y,x)$ be the space of solutions of \eqrf{instanton} satisfying
\eqrf{asymp}.
For a generic metric $g_{ij}$ (such that the gradient flow of $W$ is
of Morse-Smale type), $F_u$ is onto and hence $\ind(F_u)$ is equal to
the dimension of $\M(y,x)$.

	\SECT{3}{\mq interpretation of the topological sigma model}

Topological sigma model is an analog of supersymmetric quantum mechanics
in a more complicated situation.
Instead of the loop space, we start with the space $\map$ of maps from a
Riemann surface $\Sig$ (with complex structure $\eps$) to a symplectic manifold
$(M,\om)$ with a compatible almost complex structure $J$.\footnote{The
symplectic structure however is not important in the construction of the
Lagrangian [\ref{sig}] or in the \mq interpretation.}
Let $g$ be the induced Riemannian metric on $M$.
A natural generalization of the section $u\mapsto\dot{u}$ when $\Sig=S^1$
is $u\mapsto\dr u$, which is not a tangent vector of $\map$.
For each $u\in\map$, $\dr u$ can be regarded as a section of the bundle
$u^*TM\otimes T^*\Sig$ over $\Sig$.
So $u\mapsto\dr u$ is a section of a vector bundle $\E\to\map$ whose fiber
over $u$ is $\E_u=\Gam(u^*TM\otimes T^*\Sig)$.
Choosing local coordinates $\{\sig^\al\}$ of $\Sig$ and $\{x^i\}$ of $M$,
a local basis of $T(\map)$ is $\{\sfdo{u^i(\sig)}\}$.
A tangent field on $\map$ has the form similar to \eqrf{vect}:
	\be
\PHI=\int_\Sig\dr^2\sig\,\PHI^i(u,\sig)\fdo{}{u^i(\sig)}.
	\ee
A section of $\E$ is
	\be\eqlb{locsec}
\PSI=\int_\Sig\dr^2\sig\,\PSI^i_\al(u,\sig)\fdo{}{u^i(\sig)}\otimes\dr\sig^\al,
	\ee
or simply denoted by its components $\PSI^i_\al(u,\sig)$.
The bundle $\E$ has a natural connection.
Consider the evaluation map $\ev\colon\map\times\Sig\to M$.
Let $\pi_1$ and $\pi_2$ be the canonical projections of $\map\times\Sig$
onto $\map$ and $\Sig$, respectively.
A section of $\E$ can be canonically identified with one of
$\ev^*TM\otimes\pi^*_2T^*\Sig$,
i.e., $\Gam(\E)\cong\Gam(\ev^*TM\otimes\pi^*_2T^*\Sig)$.
In fact, the bundle $\E$ is the push-forward of $\ev^*TM\otimes\pi^*_2T^*\Sig$
via $\pi_1$.
The Levi-Civita connection on $TM$ pulls back to $\ev^*TM$.
Choosing a metric $h_{\al\beta}$ on $\Sig$ compatible with the complex
structure $\eps$, $T\Sig$ (hence $T^*\Sig$) has a Levi-Civita connection,
which pulls back to $\pi^*_2T^*\Sig$.
Thus we have a connection on $\ev^*TM\otimes\pi^*_2T^*\Sig$ by taking the
tensor product.
Finally the connection on $\E$ is obtained by restricting the covariant
derivative to the tangent directions of $\map$ in the base manifold.
A simple calculation shows
	\be\eqlb{20}
\nabla_\sfdo{u^i(\tau)}\PSI^k_\al(u,\sig)=\fdo{\PSI^k_\al(u,\sig)}{u^i(\tau)}
	+\chr{k}{ij}(u(\sig))\,\del^{(2)}(\sig-\tau)\PSI^j_\al(u,\sig),
	\ee
which is independent of the metric $h_{\al\beta}$ on $\Sig$.
After calculations similar to those to obtain \eqrf{11} and \eqrf{12},
we find that the covariant derivative of the section $u\mapsto\dr u$
along $\PHI$ is
	\be
\nabla_\PHI\pdr_\al u^i(\sig)=D_\al\PHI^i(\sig),
	\ee
where $D_\al\PHI^k=\pdr_\al\PHI^k+\chr{k}{ij}\pdr_\al u^i\PHI^j$ is given by
the pull-back connection on $u^*TM$.

The problem with the bundle $\E$ in the \mq construction is that the rank
of $\E$ is greater than the dimension of $\map$ by an infinite amount.
More precisely, the linearization
$\dr\colon\Gam(u^*TM)\to\Gam(u^*TM\otimes T^*\Sig)$ of the section
$u\mapsto\dr u$ is not a Fredholm operator.
This is resolved by restricting $\E_u$ to its anti-holomorphic part
$\EA_u=\Gam((u^*TM\otimes T^*\Sig)^{01})$, i.e., the space of sections
$\PSI$ satisfying the ``anti-$J$-linearity'' constraint
	\be\eqlb{sd}
\eab\PSI^i_\beta=-J^i_{\;\;j}\PSI^j_\al.
	\ee
The sub-bundle $\EA$ of $\E$ has a connection $\nabla^{01}$ defined
by projection, i.e.,
	\be\eqlb{proj}
\nabla^{01}\PSI^k_\al(u,\sig)=
\hf(\nabla\PSI^k_\al(u,\sig)+\eab J^k_{\;\;j}\nabla\PSI^j_\beta(u,\sig)).
	\ee
In $\EA$, there is a natural section
$u\mapsto\pdb_Ju=\hf(\dr u+J\circ\dr u\circ\eps)$,
or $\pdb_\al u^i=\hf(\pdr_\al u^i+\eab J^i_{\;\;j}\pdr_\beta u^j)$.
Solutions to the equation $\pdb_Ju=0$ are called pseudo-holomorphic (or
$J$-holomorphic) curves in $M$ [\ref{g}].
The covariant derivative of the section $u\mapsto\pdb_Ju$ along
a tangent vector $\PHI$ at $u\in\map$ is, taking into account
the variation of the almost complex structure $J$,
	\be\eqlb{21}
\nabla^{01}_\PHI\,\pdb_\al u^k=\hf(D_\al\PHI^k+\eab J^k_{\;\;i}D_\beta\PHI^i)
+\inv{4}J^k_{\;\;i;j}\PHI^j(\eab\pdr_\beta u^i+J^i_{\;\;l}\pdr_\al u^l).
	\ee
In a coordinate-free language, for every $\PHI\in\Gam(u^*TM)$,
	\be\eqlb{22}
\nabla^{01}_\PHI(\pdb_J)=\hf(D\PHI+J\circ D\PHI\circ\eps)
			+\inv{4}D_\PHI J\circ(\dr u\circ\eps+J\circ\dr u)
	\;\in\Gam((u^*TM\otimes T^*\Sig)^{01}).
	\ee

With this explicit expression, we can interpret the BRST algebra
and the Lagrangian of the topological sigma model in [\ref{sig}]\footnote{In
[\ref{bs}], it was realized that this Lagrangian can be obtained by
gauge fixing a topological action.}.
The fields consist of a (bosonic) map $u\in\map$, two fermionic fields
$\chi\in\Gam(u^*TM)$, $\rho\in\Gam((u^*TM\otimes T^*\Sig)^{01})$, and
a bosonic field $H$, a section in the same bundle as $\rho$.
So the fields $\rho^i_\al$ and $H^i_\al$ obey the same ``anti-$J$-linearity''
constraint \eqrf{sd}.
At the classical level, there is a bosonic symmetry with charge
$U=0,1,-1,0$ on $u$, $\chi$, $\rho$, $H$, respectively, corresponding to
the grading of differential forms on the moduli space.
The BRST supersymmetry is
	\bea
\del u^i\eq i\chi^i,	\quad\quad\quad\quad \del\chi^i=0,	\\
\del\rho_{\al i}\eq H_{\al i}-
   i(\chr{k}{ij}\del^\beta_\al+\hf\eab J^k_{\;\;i;j})\rho_{\beta k}\chi^j,
						\eqlb{drho}	\\
\del H_{\al i}\eq
   i(\chr{k}{ij}\del^\beta_\al+\hf\eab J^k_{\;\;i;j})H_{\beta k}\chi^j	\nno
\vc-\inv{4}(R^l_{\;\;ijk}+R^m_{\;\;\;njk}J^l_{\;\;m}J^n_{\;\;i}
            +J^l_{\;\;m:j}J^m_{\;\;\;i;k})\rho_{\al l}\chi^j\chi^k.   \eqlb{dH}
	\eea
The terms proportional to $\chi^i$ in \eqrf{drho} and \eqrf{dH}
correspond to those in \eqrf{21} that are proportional to $\dr u^j$.
Since $\del^2\rho=0$, the remaining terms in \eqrf{dH} give the curvature
$\R$ of the bundle $\EA$.
The Lagrangian can be chosen as
$\lag=\del(\rho^\al_i(\pdr_\al u^i-\inv{4}H^i_\al))$.
After eliminating $H^i_\al$ using the equation of motion
$H^i_\al=\pdr_\al u^i+\eab J^i_{\;\;j}\pdr_\beta u^j$, we get
	\bea\eqlb{lag2}
\lag\eq\hf g_{ij}\pdr_\al u^i\pdr^\al u^j
+\hf\eps^{\al\beta}J_{ij}\pdr_\al u^i\pdr_\beta u^j
-i\rho^\al_i(D_\al\chi^i+\hf\eab J^i_{\;\;j;k}\chi^k\pdr_\beta u^j)	\nno
\vc+\inv{8}(R^{ij}_{\;\;\;kl}+\hf J^i_{\;\;m;k}J^{jm}_{\;\;\;\;\;;l})
\rho_{\al i}\rho^\al_j\chi^k\chi^l.
	\eea
This is symbolically
	\be
\lag\sim\hf|\pdb u|^2-i\rho\nabla_\chi(\pdb)+\inv{4}\R\chi\chi\rho\rho,
	\ee
which agrees with the exponent of \eqrf{char}.

Similar to Section 2, \mq interpretation here is based on the result that
the partial differential operator acting on $\PHI$ on the right hand side of
\eqrf{21} or \eqrf{22} is the linearization of the section $u\mapsto\pdb_Ju$;
this linearization again appears in the Lagrangian \eqrf{lag2}.
Let $\M$ be a connected component of the moduli space of pseudo-holomorphic
curves $(\pdb_J)^{-1}(0)$.
For a generic almost complex structure $J$, the linearization of $\pdb_J$
is onto; its index is equal to the dimension of $\M$.
Applying Riemann-Roch theorem, we have [\ref{g}]
	\be
\dim\M=(1-g)\dim M+2c_1(u^*TM).
	\ee

To study the canonical formalism and Floer homology,
take $\Sig=\RE\times S^1$ with coordinate $\sig=t+is$.
$T^*\Sig$ has two global sections $\dr s$ and $\dr t$ satisfying
$\dr s\circ\eps=\dr t$ and $\dr t\circ\eps=-\dr s$.
A section $\PSI$ of $\E^{01}$ is of the form
	\be
\PSI=\hf\PSI_1\otimes\dr t-\hf J\PSI_1\otimes\dr s,
	\ee
where $\PSI_1$ is a tangent vector field of $\map$.
In other words, $\E^{01}$ and $T(\map)$ are isomorphic as bundles
(though equipped with different connections).
The section $\pdb_Ju$, for example, corresponds to the tangent vector
	\be\eqlb{cr}
(\pdb_Ju)^i_1=\spdo{u^i}{t}+J(u)\spdo{u^i}{s}.
	\ee
Naturally, \eqrf{cr}=0 is the (non-linear) Cauchy-Riemann equation for
pseudo-holomorphic curves.
One can introduce the analogue of the Morse potential in Section 2.
Let $H\colon M\times S^1\to\RE$ be a (time-dependent) Hamiltonian function
on the symplectic manifold $M$.
Pulling back via the evaluation map, the gradient of $H$ can be regarded as
a tangent vector field of $\map$, still denoted by $\grad H$.
According to the above discussion, the tangent vector
	\be\eqlb{grad}
\spdo{u}{t}+J\spdo{u}{s}+\grad H(u,s)
	\ee
defines a section of $\E^{01}$.
A direct calculation shows that the action of the covariant derivative
$\nabla^{01}_\PHI$ on \eqrf{grad} is
	\be\eqlb{33}
D_t\PHI+JD_s\PHI+\nabla_\PHI\,\grad H+
\hf\nabla_\PHI J(J\spdo{u}{t}+\spdo{u}{s}+J\,\grad H).
	\ee
In fact \eqrf{grad}=0 is the gradient flow of a Morse function on $LM$
whose critical points are periodic trajectories under the (time-dependent)
Hamiltonian flow of $H$, under certain topological assumptions on $M$
and its symplectic form $\om$.
If $u(s,t)$ does satisfies \eqrf{grad}=0, then \eqrf{33} reduces to
	\be\eqlb{34}
D_t\PHI+JD_s\PHI+\nabla_\PHI\grad H+\nabla_\PHI J\circ\spdo{u}{t}.
	\ee
Here again, \eqrf{34} is the linearization of \eqrf{grad}.
The index of the operator in \eqrf{34} can be used to associate a grading
on the set of periodic trajectories of the Hamiltonian flow [\ref{f}]
(see also [\ref{s}]).
The resulting Floer homology group is useful in solving the Arnold
conjecture [\ref{a}].

	\SECT{4}{Gauged topological sigma models}

In this section, we study the geometry of fibered Riemannian manifolds
and discuss the corresponding \mq construction and the
topological sigma model coupled to this geometric background,
which includes an important special case of coupling to gauge fields.

Let $X\to\Sig$ be a smooth fibration such that each fiber is diffeomorphic to
a manifold $M$ and is equipped with a (fiber-dependent) Riemannian metric $g$.
Under a local trivialization, $X$ can be described by the coordinates
$\{\sig^\al\}$ of $\Sig$ and $\{x^i\}$ of $M$, and $\{\pdr_\al,\pdr_i\}$
is a basis of the $TX$.
The relative tangent bundle $\rt$ of this fibration is a vector bundle
over $X$ whose fiber at each point is the tangent space to the fiber,
i.e., spanned by $\{\pdr_i\}$.
The Levi-Civita connection on the fiber defines the parallel transport of
vertical vectors along the vertical directions.
We choose a splitting of $TX$ into $\rt$ and horizontal subspaces.
Under an arbitrary local trivialization, $\pdr_\al$ is not necessarily
a horizontal vector.
Let $f^i_\al\pdr_i$ be its vertical component, then its horizontal
component is $\pdt_\al=\pdr_\al-f^i_\al\pdr_i$.
This splitting defines a connection of $\rt$, for it determines
the parallel transport of points, and hence curves in the fibers
along horizontal directions in $X$, and by differentiating, we know
how to parallel transport vertical vectors along horizontal directions.
Using the above coordinates, the covariant derivative is
$\nabla_{\pdt_\al}\pdr_j=f^i_{\al,j}\pdr_i$ or
$\nabla_{\pdr_\al}\pdr_j=f^i_{\al;j}\pdr_i$.
So the \Chr symbols $\chr{i}{\al j}=f^i_{\al;j}$.
In a coordinate-free language, if $H$ is the horizontal lift of a vector field
on $\Sig$ and $V$ is a vertical vector field on $X$, then $\nabla_HV=[H,V]$.

A section $u\colon\Sig\to X$ is locally represented by
$\sig^\al\mapsto(\sig^\al,u^i(\sig))$.
We define the covariant differential $\nabla u$ of $u$ as the projection of
$\dr u=\dr\sig^\al\otimes\pdr_\al+\pdr_\al u^i\dr\sig^\al\otimes\pdr_i$
onto the vertical directions, i.e.,
$\nabla u=(\pdr_\al u^i+f^i_\al)\dr\sig^\al\otimes\pdr_i$.
The tangent space of the space of sections $\sect$ at $u$ is $\Gam(u^*\rt)$.
Clearly, $u\mapsto\nabla u$ is a section of the bundle $\E\to\sect$,
with $\E_u=\Gam(u^*\rt\otimes T^*\Sig)$.
An arbitrary section of $\E$ locally has the same form as \eqrf{locsec}
and can be identified with one of $\ev^*\rt\otimes\pi^*_2T^*\Sig$ over
$\sect\times\Sig$.
Here $\ev\colon\sect\times\Sig\to X$ is the evaluation map and $\pi_2$
is the projection of $\sect\times\Sig$ onto $\Sig$.
Taking the tensor product of the pull-back connections from $\rt$ and
$T^*\Sig$, we get a connection on $\E$, which locally is still given
by \eqrf{20}.
Assume that $\Sig$ is a Riemann surface with complex structure $\eps$
and that there is an almost complex structure on each fiber of $X$,
i.e., a section $J$ of $\mbox{End}(\rt)$ such that $J^2=-1$.
We restrict the bundle $\E$ to its anti-holomorphic part, i.e.,
$\EA_u=\Gam((u^*\rt\otimes T^*\Sig)^{01})$, the space of sections
satisfying \eqrf{sd}.
The sub-bundle $\EA$ has a connection $\nabla^{01}$ defined by
projection \eqrf{proj}.
The natural section of $\EA$ is
$u\mapsto\nabb_Ju=\hf(\nabla u+J\circ\nabla u\circ\eps)$, or $\nabb_\al u^i
=\hf[(\pdr_\al u^i+f^i_\al)+\eab J^i_{\;\;j}(\pdr_\beta u^j+f^j_\beta)]$.
Along any tangent vector $\PHI\in T_u\sect$, the covariant derivative
$\nabla^{01}_\PHI(\nabb_J)$ is formally given by the same formula
\eqrf{21} or \eqrf{22}, but
$D_\al\PHI^k=\pdr_\al\PHI^k+(\chr{k}{ij}\pdr_\al u^i+f^k_{\al;j})\PHI^j$
is the pull-back connection on $u^*\rt$.

Consider the topological sigma model coupled to this non-dynamical
fibration $X\to\Sig$ as background, with $u\in\sect$, $\chi\in\Gam(u^*\rt)$
and $\rho,H\in\Gam((u^*\rt\otimes T^*\Sig)^{01})$.
They have the same statistics and the charge $U$ as before.
Moreover, the BRST algebra stays the same. (The terms in
$\nabla^{01}_\PHI(\nabb_J)$ proportional to $\pdr_\al u^i$ do not change.)
But the Lagrangian is replaced by
	\be
\lag=\del(\rho^\al_i(\nabla_\al u^i-\inv{4}H^i_\al)).
	\ee
After using the equation of motion
$H^i_\al=\nabla_\al u^i+\eab J^i_{\;\;j}\nabla_\beta u^j$, $\lag$ has the same
form as \eqrf{lag2}, except that $\pdr_\al u^i$ is replaced by $\nabla_\al u^i$
and that $D_\al$ is the covariant derivative on $u^*\rt$.
So this sigma model coupled to the geometry of the fibration is topological
in the sense that the Lagrangian is obtained by gauge-fixing the trivial
action,
and that the stationary phase approximation in the path integral is exact.
However, the \mq interpretation works in the conventional sense only when
the connection $\nabla^{01}$ in $\E^{01}$ is metric-preserving.
This would require that the connection $\nabla$ in $\rt$ is so,
a statement not necessarily true.
However, there is another natural connection on $\rt$ that appeared in
family index theorem [\ref{bf}, \ref{bgv}] and topological gravity [\ref{wu}].

Recall that each fiber is equipped with a metric $g_{ij}$ and that we have
chosen horizontal subspaces at each point (locally characterized by $f^i_\al$).
Together with a metric $h_{\al\beta}$ on $\Sig$, we can construct a metric
	\be
\gt=\left(	\begin{array}{cc}
		g_{ij}		&	g_{ij}f^j_\beta		\\
		f^i_\al g_{ij}	&	h_{\al\beta}+f^i_\al f^j_\beta g_{ij}
		\end{array}
    \right)
	\ee
on $X$ such that the splitting of $TX$ into vertical and horizontal subspaces
is orthogonal under $\gt$.
The projection of the Levi-Civita connection $\tilde{\nabla}$ on $X$ onto
the vertical directions defines a connection $\nabla'$ of $\rt$.
In terms of \Chr symbols, this connection is given by
	\be
\chrp{k}{ij}=\chrt{k}{ij}+\chrt{\beta}{ij}f^k_\beta
	\ee
and
	\be
\chrp{k}{\al j}=\chrt{k}{\al j}+\chrt{\beta}{\al j}f^k_\beta.
	\ee
Along the vertical directions, $\tilde{\nabla}$ agrees with the Levi-Civita
connection on the fiber, i.e., $\chrt{k}{ij}=\chr{k}{ij}$.
A more or less lengthy calculation shows that
	\be\eqlb{prime}
\chrp{k}{\al j}=f^k_{\al;j}+\hf g^{ki}(L_{\pdt_\al}g)_{ij},
	\ee
where
	\be
(L_{\pdt_\al}g)_{ij}=g_{ij,\al}
                     -(f^k_{\al,i}g_{kj}+f^k_{\al,j}g_{ik}+f^k_\al g_{ij,k})
	\ee
is the Lie derivative of $g_{ij}$ with respect to the horizontal vector
$\pdt_\al$. (Here and after, the indices $i,j,k,\dots$
are raised by the inverse $g^{ij}$ of $g_{ij}$, not by
$\gt^{ij}=g^{ij}+f^i_\al f^j_\beta h^{\al\beta}$.)
Taking into account $\chr{i}{\al j}=f^i_{\al;j}$, \eqrf{prime} agrees with
the coordinate-free expressions in [\ref{bgv}]\footnote{In [\ref{bf}],
it was shown that if there is a hermitian connection in a vector bundle
over $X$, the induced connection in the (infinite dimensional) push-forward
bundle over $\Sig$ is not necessarily so. However it could be made hermitian
by adding a term similar to the right side of \eqrf{diff} below.}.

We now compare two connections $\nabla$ and $\nabla'$ on $\rt$.
Both of them are independent of the metric $h_{\al\beta}$ on $\Sig$ and both
are equal to the Levi-Civita connection of the fiber along vertical directions.
The connection $\nabla'$ is metric preserving, but $\nabla$ is not in general.
In fact
	\be\eqlb{diff}
\chrp{k}{\al j}-\chr{k}{\al j}=\hf g^{ki}(L_{\pdt_\al}g)_{ij}.
	\ee
So $\nabla'=\nabla$ if and only if $L_{\pdt_\al}g=0$, that is, when
the parallel transport generates isometries among the fibers.
This turns out to be a very important case.
The fibration $\pi\colon X\to\Sig$ is an associated bundle of a principal
$\diff(M)$-bundle.
Over each point $\sig\in\Sig$, the fiber of this principal bundle is
$\diff(M,\pi^{-1}(\sig))$, with the right action of $\diff(M)$ by composition.
The connection of the principal bundle is defined by composing the maps in
$\diff(M,\pi^{-1}(\sig))$ with the parallel transport between the fibers.
When $L_Hg=0$ for any horizontally lifted vector field $H$, the holonomy
around any loop in $\Sig$ lies in the (finite dimensional) compact Lie group
$G$ of isometries of $M$.
So the structure group can be reduced to $G$ and $X$ is an associated bundle
of a principal $G$-bundle $P$, i.e., $X=P\times_GM$.

If we start with a principal $G$-bundle $P$ and assume that $G$ acts on $M$
preserving the metric $g$ and the almost complex structure $J$,
then the associated bundle $X=P\times_GM$ is a fibration of Riemannian
manifolds with almost complex structures on the fibers.
A connection on $P$, locally given by the gauge potential $A^a_\al$ on $\Sig$,
defines the horizontal subspaces in $TX$.
Let $V_a, a=1,\dots,\dim G$, be the Killing vector fields on $M$
induced by the Lie algebra action, then under the induced local product
structure of $X$, $f^i_\al=A^a_\al V^i_a$ and $\pdt_\al=\pdr_\al-A^a_\al V_a$.
The Lie derivatives $L_{\pdr_\al}$ and $L_{A^a_\al V_a}g=A^a_\al L_{V_a}g$
are separately zero, hence $L_{\pdt_\al}g=0$.
The Lagrangian is \eqrf{lag2} with $\pdr_\al u^i$ replaced by
$\nabla_\al u^i=\pdr_\al u^i+A^a_\al V^i_a$ and
$D_\al\chi^k=\pdr_\al\chi^k+(\chr{k}{ij}\pdr_\al u^i+A^a_\al V^k_{\al;j})$.
In this case, the connection $\nabla^{01}$ defined in the infinite
dimensional vector bundle $\E^{01}$ is metric-preserving.

The BRST invariant observables are constructed from the $G$-equivariant
differential forms on $M$ [\ref{sig}].
(The latter is related to the differential forms on the symplectic
quotient via the Kirwan map.)
Each equivariant form on $M$ can be extended to a form on $X$, pulled back
to $\sect\times\Sig$, integrated along a homology cycle in $\Sig$, and
restricted to the moduli space $\M$ in $\sect$.
It would be interesting to relate the intersection ring of $\M$ to the
cohomology ring of the symplectic quotient.\\

The author would like to thank Gang Tian and Weiping Zhang for helpful
discussions. This work is supported in part by NSF grant DMS-9305578.

        \newcommand{\rf}[1]{\item \label{#1}}

        \newcommand{\athr}[2]{{#1}.$\,${#2}}
        \newcommand{\au}[2]{\athr{{#1}}{{#2}},}
        \newcommand{\an}[2]{\athr{{#1}}{{#2}} and}

        \newcommand{\jr}[6]{{\it {#1}}, {#2} {#3} ({#4}) {#5}-{#6}}

        \newcommand{\pr}[3]{{\it {#1}}, {#2} ({#3})}

        \newcommand{\bk}[4]{{\it {#1}} ({#2}, {#3}, {#4})}

        \newcommand{\cf}[8]{{\it {#1}}, in: {\it {#2}}, {#5}, pp.$\,${#3}-{#4}
                 ({#6}, {#7}, {#8})}

\newpage
        \begin{center}
{\bf References}
        \end{center}
{\small
        \newcounter{rfs}
        \begin{list}%
        {[\arabic{rfs}]}{\usecounter{rfs}}

\rf{g}
\au{M}{Gromov}
\jr{Pseudoholomorphic curves in symplectic manifolds}
{Invent. Math.}{82}{1985}{307}{347}

\rf{mdal}
\au{D}{McDuff}
\jr{Elliptic in symplectic geometry}
{Bull. Amer. Math. Soc.}{23}{1990}{143}{16};\\
\an{M}{Audin} \athr{J}{Lafontaine} (Eds.),
\bk{Holomorphic curves in symplectic geometry}
{Birkh\"auser}{Basel, Boston, Berlin}{1994}

\rf{sig}
\au{E}{Witten}
\jr{Topological sigma models}
{Commun. Math. Phys.}{118}{1988}{411}{449}

\rf{mq}
\an{V}{Mathai} \au{D}{Quillen}
\jr{Superconnections, Thom classes, and equivariant differential forms}
{Topology}{25}{1986}{85}{110}

\rf{blau}
\au{M}{Blau}
\jr{The \mq formalism and topological field theory}
{J. Geom. Phys.}{11}{1993}{95}{127}

\rf{tym}
\au{E}{Witten}
\jr{Topological quantum field theory}
{Commun. Math. Phys.}{117}{1988}{353}{386}

\rf{aj}
\an{M.$\,$F}{Atiyah} \au{L.$\,$C}{Jeffrey}
\jr{Topological Lagrangians and cohomology}
{J. Geom. Phys.}{7}{1990}{119}{136}.

\rf{sqm}
\au{E}{Witten}
\jr{Dynamical breaking of supersymmetry}
{Nucl. Phys.}{B185}{1981}{513}{554};
\jr{Constraints on supersymmetry breaking}
{Nucl. Phys.}{B202}{1982}{253}{316};
\jr{Supersymmetry and Morse theory}
{J. Diff. Geom.}{17}{1982}{661}{692}

\rf{k}
\au{W}{Klingenberg}
\bk{Riemannian geometry}
{Walter de Gruyter}{Berlin, New York}{1982}, chap.$\,$1

\rf{bs}
\an{L}{Baulieu} \au{I.$\,$M}{Singer}
\jr{The topological sigma model}
{Commun. Math. Phys.}{125}{1989}{227}{237}

\rf{s}
\au{D}{Salamon}
\jr{Morse theory, the Conley index and Floer homology}
{Bull. London Math. Soc}{22}{1990}{113}{140}

\rf{f}
\au{A}{Floer}
\jr{Symplectic fixed point and holomorphic spheres}
{Commun. Math. Phys.}{120}{1989}{575}{611}

\rf{a}
\au{V.$\,$I}{Arnold}
\jr{Sur une propriet\'e topologique des applications globalement canoniques
de la m\'eca- nique classique}{C. R. Acad. Sci. Paris}{261}{1965}{3719}{3722};
\jr{First steps in symplectic geometry}
{Russian Math. Surveys}{41}{1986}{1}{21}

\rf{bf}
\au{J.-M}{Bismut}
\jr{The index theorem for families of Dirac operators: two heat equation
proofs}{Invent. Math.}{83}{1986}{91}{151};\\
\an{J.-M}{Bismut} \au{D.$\,$S}{Freed}
\jr{The analysis of elliptic families II. Dirac operators, \^eta invariants,
and the holonomy theorem}{Commun. Math. Phys.}{107}{1986}{103}{163}

\rf{bgv}
\au{N}{Berline} \an{E}{Getzler} \au{M}{Vergne}
\bk{Heat kernels and Dirac operators}
{Springer-Verlag}{Berlin, New York}{1992}, chap.$\,$10

\rf{wu}
\au{S}{Wu}
\jr{Appearance of universal bundle structure in four dimensional topological
gravity}{J. Geom. Phys.}{12}{1993}{205}{215}

	\end{list}}
        \end{document}